\documentclass[conference]{IEEEtran}
\ifCLASSINFOpdf
\else
\fi
\usepackage[noend]{algpseudocode}
\usepackage{algorithmicx,algorithm}
\usepackage{cite}
\usepackage{multirow}
\usepackage{amsmath}  
\usepackage{float}
\usepackage{booktabs} 

\usepackage{graphicx}
\usepackage{subfigure}
\usepackage{bm}
\usepackage{lineno}
\usepackage{url}
\usepackage{tikz}
\usepackage{verbatim}
\usetikzlibrary{positioning}
\usepackage{calc}
\usepackage{amsfonts}

\def\BibTeX{{\rm B\kern-.05em{\sc i\kern-.025em b}\kern-.08em T\kern-.1667em\lower.7ex\hbox{E}\kern-.125emX}}

\begin{document}
\flushbottom
%
\title{Hierarchical Representation Network  for  Steganalysis of QIM Steganography in Low-Bit-Rate Speech Signals}

\author{\IEEEauthorblockN{Hao Yang}
\IEEEauthorblockA{\textit{Department of Electronic Engineering} \\
\textit{Tsinghua University}\\
Beijing, China \\
yanghao17@mails.tsinghua.edu.cn}

\and
\IEEEauthorblockN{ZhongLiang Yang}
\IEEEauthorblockA{\textit{Department of Electronic Engineering} \\
\textit{Tsinghua University}\\
Beijing, China \\
yangzl15@mails.tsinghua.edu.cn}


\and
\IEEEauthorblockN{YongFeng Huang}
\IEEEauthorblockA{\textit{Department of Electronic Engineering} \\
\textit{Tsinghua University}\\
Beijing, China \\
yfhuang@tsinghua.edu.cn}
\and
\thanks{H.Yang, Z.Yang, Y. Bao and Y. Huang are with the Department of Electronic Engineering, Tsinghua University, Beijing, 100084, China. E-mail: yanghao17@mails.tsinghua.edu.cn}
}

\maketitle

\begin{abstract}
With the Volume of Voice over IP (VoIP) traffic rises shapely, more and more VoIP-based steganography methods have emerged in recent years, which poses a great threat to the security of cyberspace. Low bit-rate speech codecs are widely used in the VoIP application due to its powerful compression capability. QIM steganography makes it possible to hide secret information in VoIP streams.  Previous research mostly focus on capturing the inter-frame correlation or inner-frame correlation features in code-words but ignore the hierarchical structure which exists in speech frame. In this paper, motivated by the complex multi-scale structure, we design a Hierarchical Representation Network to tackle the steganalysis of QIM steganography in low-bit-rate speech signal. In the proposed model, Convolution Neural Network (CNN) is used to model the hierarchical structure in the speech frame, and three level of attention mechanisms are applied at  different convolution block, enabling it to attend differentially to more and less important content in speech frame. Experiments demonstrated that the steganalysis performance of the proposed method can outperforms the state-of-the-art methods especially in detecting both short and low embeded speech samples. Moreover, our model needs less computation and has higher time efficiency to be applied to real online services.

\end{abstract}


%
\IEEEpeerreviewmaketitle

\section{Introduction}
Steganalysis and steganography are  two different sides of the same coin. Steganography tries to hide messages in plain sight while steganalysis tries to detect their existence or even more to retrieve the embedded data from suspicious carriers. In recent years, the fast growth of Internet services provides a multimedia transfer which can share enormous volumes of data over the Internet. VoIP enables the
digitalisation, compression and transmission of analogue audio signals from a sender to a receiver using IP packets. Enormous network traffic makes it suitable for steganography \cite{3-Tian2009An,4-Xu2011Adaptive,5-Dora2012Highly,6-Huang2011Steganography,7-Tian2014Improving,8-Huang2012Steganography}. Due to its real-time and large-scale characteristics, more and more VoIP-based covert communication systems were brought up in recent years \cite{VOIP-Yang2017A,6-Huang2011Steganography,Mazurczyk2013VoIP}.  This type of covert communication has become a major threat to security monitoring of network communication. Thus, it is important to develop a powerful steganalysis tool to analyze VoIP streams.

Streaming media means a continuous transaction of information, i.e., all data is not needed before receiver can take part of the information. For a long time, information hiding has focused on carriers like image and audio. A problem with these carriers is that they do not support hiding in new types of network-based service because these carriers always occur as static manner. By introducing streaming media as carrier of hidden information, hiding in new network-based services is supported. Besides,
streaming technology makes the information embedding process more dynamic and higher capacity.  
Property of information hiding technique combined with streaming media make it more challange for steganalysis compared with traditional static carrier such as iamge, text and audio.

VoIP is a typical streaming media technology. In general, VoIP streams are dynamic chunks of a series of
packets that consist of IP headers, UDP headers, RTP
headers and numbers of carrier frames. All of these fields can be used to embed secret information. However, information hiding based on network protocols including IP, UDP and RTP fields can be easily detected since all of the protocols are public and data in these fields are fixed under most conditions \cite{field-Pelaez2009Using}. On the contrary, embedding information into carrier data field or payload filed which varies with time can achieve a  relatively high level of concealment and makes them hard to detect. Low bit-rate speech coding algorithms which have powerful compression capability such as G.729 and G.723.1  standard are specially defined  by the International Telecommunication Union (ITU) for VoIP and are widely used to compress speech segment in VoIP streams. It tries to minimize the decoding error by Analysis-by-Synthesis (AbS) framework and can achieve high compression ratio while preserving superb voice quality \cite{Goode2002Voice}.

Data hiding techniques in low bit-rate speech streams can be divided into three categories according to different embedding positions. The first category embeds the secret information by revising the  value of some coded elements in the compressed speech stream. Some famous data hiding methods like Least Significant Bit (LSB) replacement methods  can be applied in this case \cite{LSB2006G}. Because Abs linear predictive coding technique is used in low bit-rate speech stream, compressed speech stream has few redundant data,  making it hard to find a proper position for data hiding. However, the second and third categories hide the secret data during the encoding process. The second category methods hide the information in the prediction step of the short term predictor (STP) of the speech codec, and data hiding methods like pitch modulation steganography  can be applied to embed data in STP step when the encoder estimates the pitch of the speech sub-frame \cite{Janicki2016Pitch}. The third category hide the information in the prediction step of the long term predictor (LTP). Representative steganography schemes like QIM steganography can be used in this step. The QIM steganography \cite{14-Chen2002Quantization} achieves information embedding by modifying the Vector  Quantization (VQ) codeword search range in the analysis step, where the introduced distortion can be compensated in the following synthesis steps of the speech encoder. The QIM method has less additional distortion is a is very common steganography scheme which can offer higher concealment capability and better robustness.


Previous research on steganalysis of QIM-based steganography in low-bit-rate speech always focus on inter-frame correlation or inner-frame correlation features but neglect the  hierarchical  structure  which  exists  in  speech  frame. In this paper, we proposed an end-to-end model  which try to model complex multi-scale structure for steganalysis of QIM-based steganography in low-bit-rate speech signals . Many natural sequences such as language, handwriting and speech have the capacity to recursively combine smaller units into hierarchically organized larger ones which is a fundamental property\cite{Hiracher-HMM-Fine1998The}. For example, in speech sequence, the phoneme, a basic phonology unit, can make up sub-words and words are composed of sub-words. The information contained in each acoustic unit is limited, but their combination leads to expressions that can flexibly convey infinite nuances and meanings. Having noticed that all the previous methods have neglected this property of speech, we try to construct a model to capture these hierachical features for steganalysis of the QIM steganography because QIM steganography can bring slight distortion to the hierarchical structure in speech. In the proposed model, CNN, regarded as a proper architecture to model hierachical structure, is stacked to capture different levels of features \cite{stackvisual}. The attention mechanism \cite{att-Vaswani2017Attention}  is used after every convolution blocks to select important components. All of the features selected from different levels of convolution blocks are concatenated and fed into fully connected layers which will serve as a classifier to indicate whether the sample speech is `stego' or `cover'. Experiments show that our model can effectively achieve the state-of-art results in both low and short samples which are the hardest parts in detecting QIM steganography in VoIP streams. Moreover, our models need less computation and has higher time efficiency to be applied to real online services.

\textbf{We summarize our main contributions as follow:}
\begin{itemize}
\item[1)] We first pointed out that speech steganalysis can make full use of the semantic hierarchical structure of speech itself, and design a reasonable network structure to model this hierarchical structure in speech carriers.
\item[2)] The end-to-end model we proposed has two distinctive characteristics: (i) It used a convolution network to model the hierarchical structure in the speech carrier, which mirrors the hierarchical structure of speech; (ii) it has different levels  of attention mechanisms applied at different level of features, enabling it to attend differentially to more and less steganalysis content when constructing the speech representation for classification.
\item[3)] Experiment on public dataset shows the proposed method can out-perform all the state-of-the-art methods espically in low embeded and short samples. Meanwhile, time efficiency of the proposed model is also excellent compared with other methods.
\end{itemize}

The rest of this paper is organized as follows. In Section \uppercase\expandafter{\romannumeral2}, we introduce some background knowledge of the research. Section \uppercase\expandafter{\romannumeral3} summarizes  the related work. In Section \uppercase\expandafter{\romannumeral4}, we introduce and describe  the details of the proposed hierarchical representation network architecture. In Section \uppercase\expandafter{\romannumeral5}, we introduce the experiment setting and benchmark. The experimental results and models are also discussed in this part. In Section \uppercase\expandafter{\romannumeral6}, the concluding remarks are given.



\section{PRELIMINARIES}
\subsection{Linear Predictive Coding}
Linear predictive coding (LPC) is a tool used mostly in speech processing for representing the spectral envelope of a digital signal of speech in compressed form, using the information of a linear predictive model. It is very useful methods for encoding good quality speech at a low bit rate and provides extremely accurate estimates of speech parameters. Speech codecs such as G.729 and G.723.1 are based on the linear
predictive coding (LPC) model, which uses an LPC filter to analyze and synthesize acoustic signals between the encoding and decoding endpoints. LPC filter can expressed as follow:
\begin{equation}
\label{LPC}
{\text{H(z) = }}\frac{1}{{A(z)}} = \frac{1}{{1 - \sum\nolimits_{i = 1}^n {{a_i}{z^{ - i}}} }},
\end{equation}
where $a_i$ is the $i$-th coefficient of the LPC filter. The short-time stationary nature of the voice signal requires the entire signal sample to be divided into frames and the LPC filter's coefficients are then computed for each frame. During the  speech coding, the LPC filter's coefficients of each frame are first computed and converted to line spectrum frequency (LSF) coefficients. Subsequently, the LSF coefficients are encoded by using vector
quantization (VQ). Speech codecs adopt split VQ and use
different split vectors to quantify the LSF coefficients and then Quantization Index Sequence (QIS) is generated, which can be formulated as:
\begin{equation}
\label{qismatrix}
S =  [{s_1},{s_2}, \cdots ,{s_T}] = \left[ {\begin{array}{*{20}{c}}
  {{s_{1,1}}}&{{s_{2,1}}} \\ 
  {{s_{1,2}}}&{{s_{2,2}}} \\ 
  {{s_{1,3}}}&{{s_{2,3}}} 
\end{array}\begin{array}{*{20}{c}}
  {.....\begin{array}{*{20}{c}}
  {} 
\end{array}\begin{array}{*{20}{c}}
  {{s_{T,1}}} \\ 
  {{s_{T,2}}} \\ 
  {{s_{T,3}}} 
\end{array}} 
\end{array}} \right]
\end{equation}
where $T$ is the total frame numbers in the sample window of the speech, $s_i$ denotes the vector in i-th frame of the speech segment, and $s_{i,j}$ denotes the j-th code-word in the i-th frame respectively.
\subsection{QIM-based steganography}
Quantization index modulation  techniques have been gaining popularity in the data hiding community because of their robustness and information-theoretic optimality against a large class of attacks.
The QIM-based VoIP steganography hides the secret data during the VQ process by embedding information in the choice of quantizers \cite{Chen2002Quantization}. For example. if we want to embed bit stream, a standard scalar QIM with two sub-codebooks $L_1$ and $L_2$ can be simply expressed as follow:
\[{s_i} = {Q_m}({x_i}) = \left\{ \begin{array}{l}
{Q_0}({x_i}){\quad}if{\ }{m_i} = 0,\\
{Q_1}({x_i}){\quad}if{\ }{m_i} = 1.
\end{array} \right.\]
$x_i$ represents the input signal, and $m$ is the message bit we want to embedded. $Q_i$ is the quantizers which choose quantitative vector from sub-codebook of $L_i$. $L_i$ is the sub-codebook of $L$ in VQ process. For a two division of codebook $L$, the sub-codebook should needs
to satisfy the following conditions:
\begin{equation}
    L1 \cap L2 = \emptyset  \quad and  \quad {\rm{ }}L1 \cup L2 = L.
\end{equation}
The receiver can recover the secret information by judging to which sub-codebook the quantitative vector belongs.


 For VoIP frames, QIM steganography are used in quantify the LSF coefficients. Obviously, QIM  steganography will have an impact on the elements of QIS. Thus, QIS is a proper clue for steganalysis of QIM steganography. Another advantage of using QIS is that we can conduct steganalysis directly in the compressed domain, which will have little impact on the users of VoIP service.

\section{RELATED WORK}
In this section, we introduce conventional steganalysis method in VoIP and deep learning based models in this field. 
\subsection{Conventional Steganalysis Method in VoIP}
Conventional steganalysis method always focus on  extracting statistical features. For example, there are some audio steganalysis methods that can be utilized for detecting the speech QIM-based VoIP steganography by extracting statistical features in the uncompressed domain \cite{Huang2011Detection, Liu2009Temporal, Kraetzer2007Mel, Kocal2008Chaotic}.  Nevertheless, these methods are not effective in detecting QIM steganography VoIP streams  which are integrated with low bit-rate speech codecs. The reason is that the method introduce minimal additional distortion in decoded speech signals. Thus it is difficult to obtain features in uncompressed domain for steganalysis. Besides, some researchers try to conduct steganalysis in the compressed domain, where the statistical characteristics of elements can be distorted during QIM steganography in speech encoding process. Therefore, the corresponding steganalysis methods usually exploit the statistical characteristics of the carrier, such as Mel-frequency features \cite{Kraetzer2007Mel}, statistic feature \cite{Huang2011Detection}, codewords correlations \cite{27-Li2017Steganalysis} and so on by manual construction to exploit the difference of statistical distribution of these features before and after steganography. Then the model can determine whether the inputted VoIP speech contains hidden information. Most of these traditional methods either have low accuracy or require a lot of computation to extract features. For example, Li $et\ al.$ \cite{27-Li2017Steganalysis} extracted the modified codewords into a data stream, and used markov chain to model the transition pattern between
successive codewords which was very time consuming.

\subsection{Deep Learning Based steganalysis Method in VoIP}
Deep learning techniques have been well applied in image \cite{He2015Deep} and natural language processing \cite{Devlin2018BERT}. Application of deep learning techniques in the field of steganalysis has also be further explored \cite{csw_yang2019realtime}. In the steganalysis of audio. C. Paulin $et\ al.$ \cite{Palaz2015Convolutional}  presents a steganalysis method that used a deep belief network (DBN) as a classifier for audio files. In another work, Paulin $et\ al.$ \cite{Paulin2016Speech}
presented a new method to train Restricted Boltzmann Machines (RBMs) using Evolutionary Algorithms (EAs), where RBMs are used in the first step of a steganalysis tool for audio files and the vector they used to train the model was MFCC.  S. Rekik $et\ al.$  \cite{23-Rekik2012An}  advocated a powerful and sophisticated classifier called Autoregressive Time Delay Neural Network (AR-TDNN). The approach uses LSF (line spectral frequencies) parameters as a cue of audio type. Wang $et\ al.$ \cite{25-Wang2019Cnn} presents an effective steganalytic scheme based on CNN for detecting MP3 steganography in the entropy code domain. The above all focused on static audio file and can't be directly applied to stream media carrier.

There are also several attempts to apply deep learning method to steganalysis  of VoIP. Lin $et\ al.$ \cite{28-Lin2018RNN} found there are four strong codeword correlation
patterns in VoIP streams, which will be distorted after embedding
with hidden data. Thus, to extract those correlation features,
they propose the codeword correlation model, which is based
on recurrent neural network (RNN). Yang $et\ al.$ \cite{csw_yang2019realtime} defined multi-channel sliding detection windows to extract feature from raw speech stream. Then, they used two feature extraction channels with CNN to extract correlations features of the input signal between neighborhood frames. The method they proposed can achieve almost real-time detection of VoIP speech signals. Although the above methods had significantly improved the performance of VoIP steganalysis, they all neglected the hierarchical structure in speech carrier which have great potential to improve the performance of steganalysis. 



\section{METHODOLOGY}
\begin{figure*}
\centering
\includegraphics[width=6in]{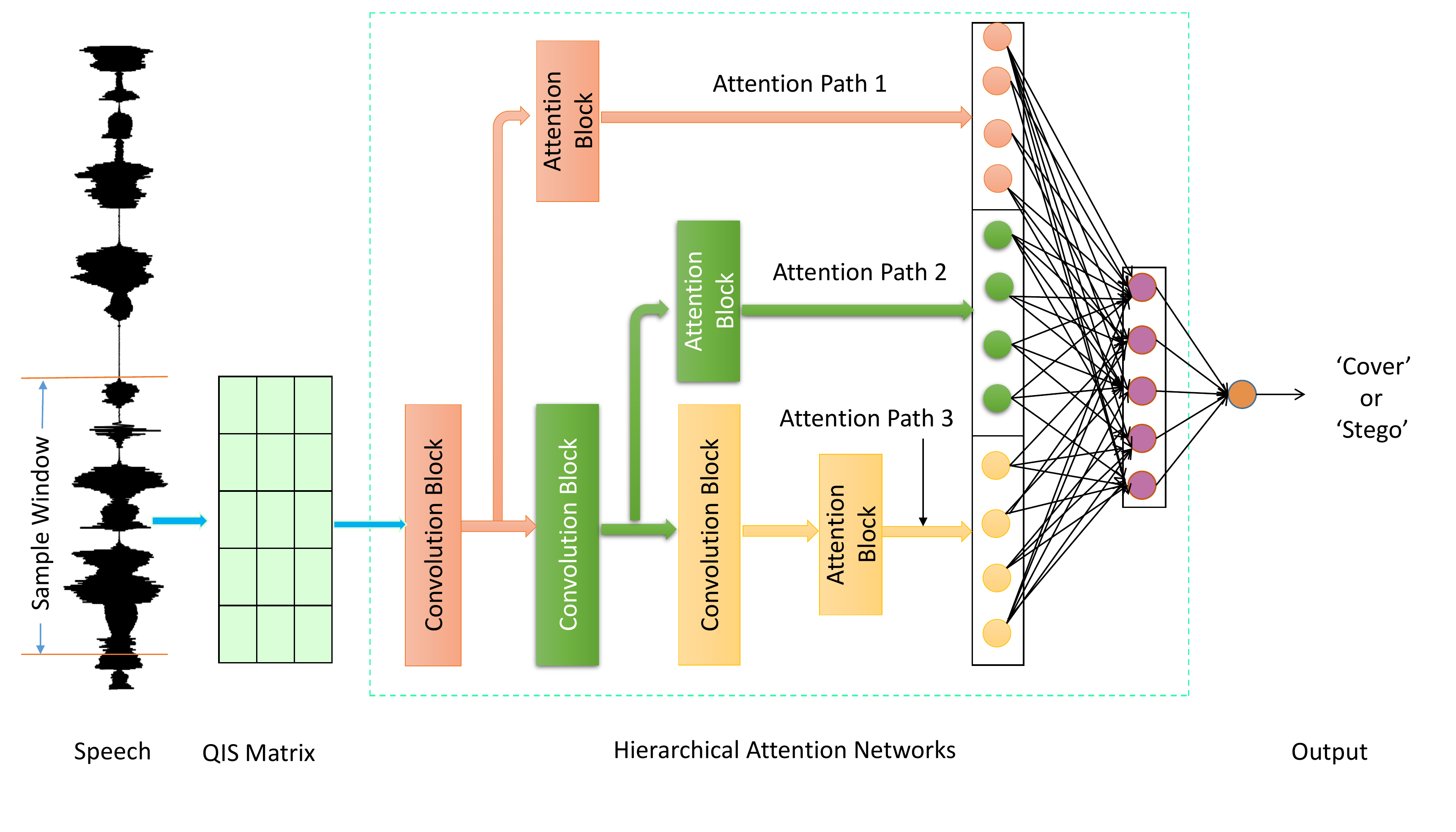}
\caption{Structure of Proposed Hierarchical Representation Network}
\label{structure}
\end{figure*}

\subsection{Problem Definition}
Steganalysis of speech streams in this paper is to judge  whether there were extra information embeded in the raw speech frame. For the online real-time speech service network system, it is unlikely to get a complete voice sample, because it will seriously affect the quality of network voice services. In general, we can only use a small window to sample a small segment of the network voice stream as our test sample. Assume that the sample window size is N, the corresponding speech sequence can be written as ${S^t} = [s_1^t,s_2^t,...,s_N^t]$, where $S_i^t$ represents speech frame code-words at time step t.   
The label for a sequence is denoted as y, $1 \le y \le C$. Our goal is to construct an end-to-end model $\phi (S^t)$ to predict a label ${\overset{\sim}y}$ 

\subsection{Model Structure}
The architecture of proposed model is shown in Figure \ref{structure}. In the proposed model, raw speech was first sampled by a sliding window [cite] and QIM sequences generated after this step. Then, the sequence were fed into the proposed model. In the proposed model, there are two main parts including feature extraction module, feature fusion and classification module.  In feature extraction module, convolution layers were cascaded to model the hierarchical structure, and attention mechanism \cite{att-Vaswani2017Attention} is used to select important features from different level. In feature fusion and classification module, features from different level are concatenated for final classification and two fully connected layers  serve as classification part in these part. Moreover, parameters in the proposed model are trained on a supervised learning framework. In the following part, we will introduce the detail parts of each module.

\subsection{Convolution Layers}
Convolution layers are the backbone of the feature extraction module. We cascade three convolution blocks to capture different levels of features. Convolution layers in our  model all used one-dimensional convolution [cite]. In the module, a filter $m$ convolves with the window vectors at each position in a valid way to generate a feature map $h$, each element $h_i$ of
the feature map for window vector $h_j$ is produced
as follows:
\begin{equation}
    {h_j} = f({s_{j:j + k - 1}} \odot m + b),
\end{equation}
where ${s_{j:j + k -1}}$ means a vector with $k$ consecutive frame vector in $S$, $\odot$ is element-wise multiplication, $b$ is a
bias term and $f$ is a nonlinear transformation function where ReLU \cite{reluNair2010Rectified} is used in our model. The structure of second and third convolution blocks is essentially similar to the first one but with different convolution kernel sizes.

\subsection{Attention Mechanism}
Attention mechanisms which can  compelling sequence modeling and transduction models have become an integral part in various tasks such as image [], NLP[], and speech[]. In the proposed model, attention mechanism are used for selecting different feature in each layer. It is generally believed that the more the neural network layers in a model, the more abstract the features will be extracted [cite]. In the feature extraction module, we use a 3-layer convolutional layer to extract hierarchical features. For steganalysis, the impact of steganography on speech stream may occur at each level of the hierarchical structure. Therefore, we believe that feature in each layer of the model are useful and they are all used for final classification. However, in each steganography sample, the importance of different levels are not equal. Thus, we introduce attention mechanism in our model to select important feature in each sample.
In the attention block of figure \ref{structure}, inputted data which generated by each convolution block can be denoted as $h = [{h_1},{h_2}, \cdots ,{h_j}]$, and for ${h_i} \in h$, its attention weight $\alpha _i$ can be formulated as follows:	
\begin{equation}
\begin{array}{l}
{m_{\rm{i}}} = \tanh ({h_i}) ,\\
\mathop {{\alpha _i}}\limits^ \wedge   = {w_i}{m_i} + {b_i} ,\\
{\alpha _i} = \frac{{\exp (\mathop {{\alpha _i}}\limits^ \wedge  )}}{{\sum\nolimits_j {\exp (\mathop {{\alpha _i}}\limits^ \wedge  )}}},
\end{array}
\end{equation}
where $w$ and $b$ are the parameters of the attention
layer. Therefore, the output representation $r$ in every attention path is given by:
\begin{equation}
  r = \sum\nolimits_i {{\alpha _i}{h_i}}.
\end{equation}

Based on such transformation, the features from convolution layers will be assigned with different attention weights. Thus, important information can be identified more easily.

\subsection{Feature Fusion and Classiﬁcation layer}
After the sample has been processed by the feature extraction module, we will get features from different levels. These features were concatenated for final classification and the compound feature vector $z$ can be denoted as follow:
\[z = [{r_1},{r_2},{r_3}]\]
 where ${r_i}$ is the representative feature from the $i$-th attention block. Generally, the dimension of $z$ is still very high, which is under the risk of over-ﬁtting. Therefore, we take a two-layer fully-connected layers to compress it. The compress process can be expressed as follow:
 \[{z_c} = {f_2}({w_2}{f_1}({w_1}z + {b_1}) + {b_2})\]
 where $w_i$ and $b_i$ is the parameters of the $i$-th fully-connected layer. $f_i$ is the activation function in $i$-th fully-connected layer and we user ReLu [cite] in our model.  
 The compressed feature vector $z_c$ is then sent to a softmax classiﬁer to  generate the probability distribution over the label set $Y$. the soft-max classiﬁer can be denoted as:
 
$${p_t}(i) = \frac{{\exp (w{}_i \cdot {z_c}{\rm{ + }}{{\rm{b}}_i})}}{{\sum\limits_{k = 1}^C {\exp ({w_k} \cdot {z_c}{\rm{ + }}{{\rm{b}}_k})} }}$$

where ${p_t}(i)$ is the probability of the category $i$ at time step $t$, the total category number is $C$. ${{w_k}}$ and ${{b_k}}$ are the  parameters in soft-max classifier. After these step,  we can get predicted label ${\overset{\sim}y}$,  which is the element position with the maximum probability in the distribution ${p_t}$ and the label value decided the speech sample belongs to 'Cover' or 'Stego'. 

\subsection{Loss Function}
The whole proposed model is trained under a supervised learning framework where cross entropy error loss is chosen as loss function of the network. Given a training sample  ${s^i}$ and its true label ${y^i} \in \{ 1,2,...,k\} $ where $k$ is the number of possible labels and the estimated probabilities ${ {\overset{\sim}y}_j^i} \in [0,1]$ for each label $ j \in \{1,2,...,k\}$, the error is defined as:
\begin{equation}
L({s^i},{y^i}) = \sum\limits_{j = 1}^k {1\{ {y^i} = j\} \log ({\overset{\sim}y}_j^i)}
\end{equation}
where 1\{condition\} is an indicator such that 1\{condition is true\} = 1 otherwise
1\{condition is false\} = 0. Moreover, in order to mitigate overfitting, we apply dropout technique \cite{Srivastava2014Dropout} and Batch Normalization \cite{bn-ref} to regularize our model.

\section{Model Evaluation and Discussion}
\begin{table*}[!htbp]
\renewcommand\arraystretch{1.3}
  \centering
  \caption{Detection Accuracy of 10s Samples Under Different Embedding Rate}
  \resizebox{6in}{!}{
    \begin{tabular}{c|l|cccccccccc}
    \toprule[2pt]
      \multirow{2}{*}{Language} &\multirow{2}{*}{Method}  &\multicolumn{10}{|c}{Embedding Rate}\\
         &  &10\% &20\% &30\% &40\% &50\% &60\% &70\% &80\% &90\% &100\%\\
    \hline
    \multirow{5}{*}{EN} & IDC \cite{Huang2011Detection} & 51.60 & 58.55 & 63.65 & 71.50 & 76.25 & 83.50 & 87.25 & 91.60 & 95.55 & 97.20 \\
       & QCCN \cite{27-Li2017Steganalysis} & 54.40 & 75.45 & 92.45 & 97.35 & 99.15 & 99.60 &  {\textbf{100.00}} &  {\textbf{100.00}} & 99.95 &  {\textbf{99.30}} \\
       & RNN-SM \cite{28-Lin2018RNN} & 59.64 & 92.44 & 94.56 & 96.90 & 97.76 & 98.77 & 99.24 & 99.71 & 99.79 & 98.78 \\
       & CSW \cite{csw_yang2019realtime} & 83.48 & 94.15 & 97.76 & 99.17 & 99.71 & 99.91 & 99.95 & 99.98 &  {\textbf{100.00}} & 99.05 \\
       & ours &  {\textbf{86.83}} &  {\textbf{95.08}} &  {\textbf{98.25}} &  {\textbf{99.53}} &  {\textbf{99.84}} &  {\textbf{99.95}} &  {\textbf{99.99}} &  {\textbf{100.00}} &  {\textbf{100.00}} & 99.13 \\
    \hline
    \hline
    \multirow{5}{*}{CH} & IDC\cite{Huang2011Detection} & 52.75 & 59.25 & 65.55 & 71.40 & 78.50 & 82.60 & 89.15 & 93.60 & 96.05 & 98.05 \\
       & QCCN \cite{27-Li2017Steganalysis} & 57.35 & 75.00 & 92.00 & 98.25 & 99.50 & 99.85 &  {\textbf{100.00}} & 99.95 & 99.90 &  {\textbf{99.75}} \\
       & RNN-SM \cite{28-Lin2018RNN} & 55.14 & 74.19 & 90.12 & 95.24 & 98.05 & 98.25 & 99.09 & 99.51 & 99.76 & 99.55 \\
       & CSW \cite{csw_yang2019realtime} & 77.18 & 92.05 & 96.58 & 98.70 & 99.64 & 99.87 & 99.94 & 99.98 &  {\textbf{100.00}} & 99.51 \\
       & ours &  {\textbf{86.54}} &  {\textbf{95.24}} &  {\textbf{98.28}} &  {\textbf{99.38}} &  {\textbf{99.81}} &  {\textbf{99.92}} &  {\textbf{100.00}} &  {\textbf{100.00}} &  {\textbf{100.00}} & 99.61 \\
    \bottomrule[2pt]
    \end{tabular}%
    }
  \label{tab:embed}%
\end{table*}%

\subsection{Dataset}
Our experiments were conducted in a public dataset  \footnote{https://github.com/fjxmlzn/RNN-SM} that has been published by Lin $et\ al$. \cite{28-Lin2018RNN}. Samples in this dataset have different types of native speakers. Each speech file in the
datasets was encoded according to the G.729a standard. Speech clips without hidden information were assigned the category label `cover' which made up the cover speech dataset, while, secret data were embedded using CNV-QIM \cite{31-Xiao2008An} steganography in split vector quantization process. Those speech samples with hiding data were assigned the category label `stego' and make up the stego speech dataset. When we conducted experiment, samples in cover speech dataset and stego speech dataset were cut into different lengths to test the model performance with different duration. Segments of the same length were successive and noverlapped. For the training set with 0.1s clips, there were 2,486,708 samples with the 1:1 ratio of cover clips and stego clips. Both the testing set and the validation set  contain 155,405 clips.

\subsection{Experimental Setting}
\subsubsection{Baselines}

In order to validate the effectiveness
of the proposed model, we compared the performance of
our model with several baseline methods. Methods to be compared include:

\noindent {\bfseries IDC} \cite{Huang2011Detection}: this method tries to exploit the Index Distribution Characteristics (IDC). The model extracted vector variation rate to measure the change of a vector and  used first-order Markov chain for quantifying the correlated features. Then, they used Support Vector Machine (SVM) for classification .

\noindent {\bfseries QCCN} \cite{27-Li2017Steganalysis}: the authors constructed a model called the Quantization code-word correlation network (QCCN) based on split VQ code-words  from adjacent speech frames. They used high order Markov to model correlation characteristics of split VQ code-words and they also used SVM for classification.

\noindent {\bfseries RNN-SM}\cite{28-Lin2018RNN}: this method indicated four strong code-word correlation patterns in VoIP streams, which will be distorted after embedding with hidden data. To extract those correlation features, the author proposed the codeword correlation model, which was based on Recurrent Neural Network (RNN).

\noindent {\bfseries CSW} \cite{csw_yang2019realtime}: in order to exploit the correlations between frames and different neighborhood frames in a VoIP signal, the method combines sliding windows and convolution neural network to conduct steganalysis in compressed domains.  

\subsubsection{Setting of the Proposed Model}
The hyperparameters in our model were selected via cross-validation on the trail set. More specifically, the convolution kernel sizes of CNN filters were 1, 3, 5 from first convolution block to the third convolution block. The number of each CNN filter in each convolution block was 256.  The dimension of fully connected layer was 64, and the dropout rate was 0.6 for fully connected layer. The batch size in training process was 256, and the maximal training epoch was set to 200 which was large enough for convergence of all the models . We used Adam \cite{adam-ref} as the optimizer for network training. Our model was implemented by Keras. 
We train all networks on GeForce GTX 1080 GPU with 16G graphics memory. Prediction process is done both on previous GPU and on ”Intel(R) Xeon(R) CPU E5-2683 v3 \@
2.00GHz”.

\subsubsection{Evalution Metrics}
The metric we chose to validate our model performance was  classification accuracy, which was defined as the ratio of the number of samples that were correctly classified to the total number of samples. 


\subsection{Evaluation Results and Discussion}

\subsubsection{Influence of Embeding rate}
\begin{figure}
\centering
\includegraphics[width=3.5in]{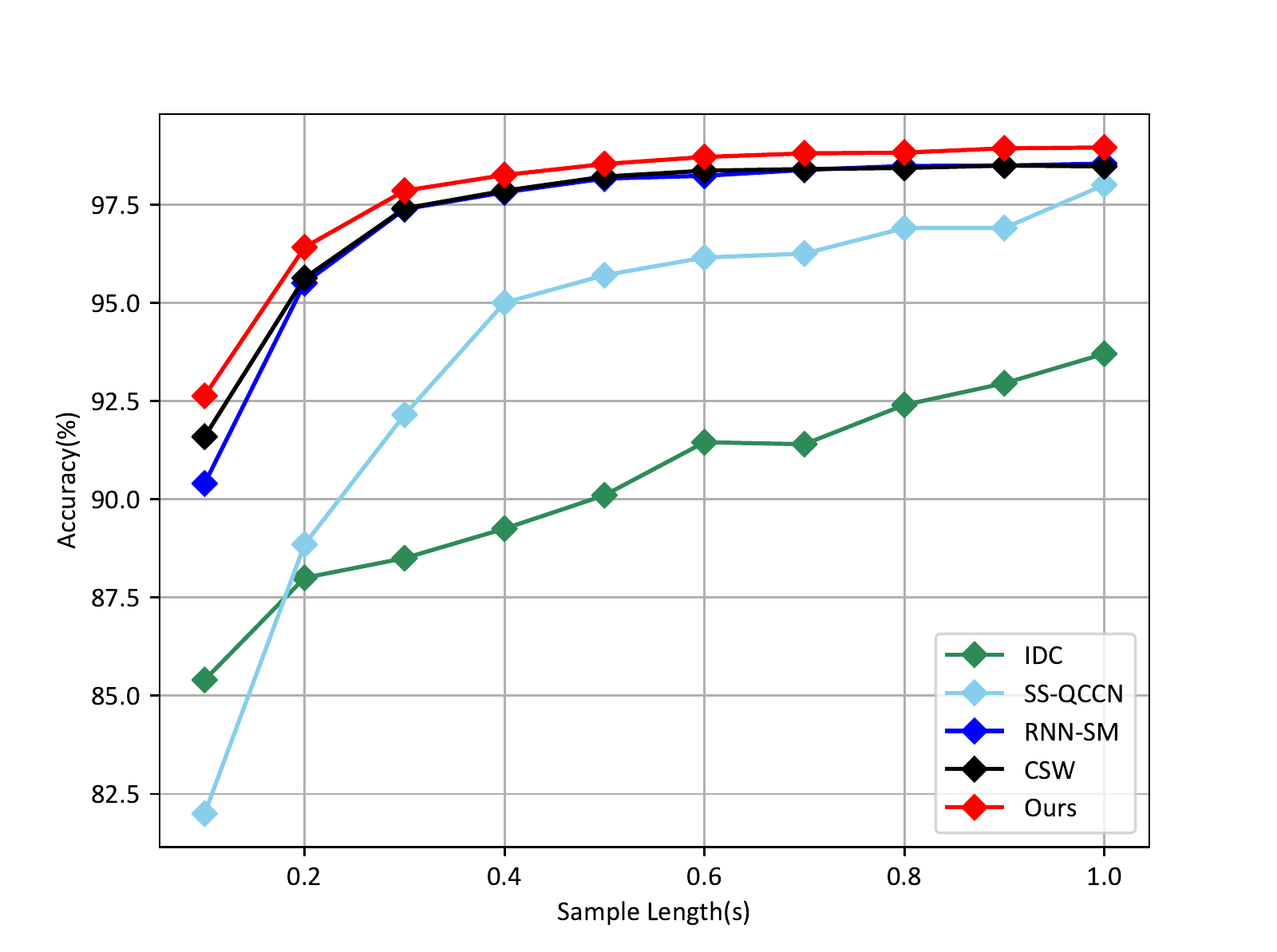}
\caption{Detection Accuracy of 10s Samples Under Different Embedding Rate}
\label{fig_embed}
\end{figure}

The embedding rate is an important factor influencing detecting accuracy. In this part, we fixed the sample length at 10s, and changed embedding rate from 10\% to 100\% with step size of 10\% to test different models. English and Chinese speeches were tested separately. The experiment results are shown in Table \ref{tab:embed} and Figure \ref{fig_embed}. As Figure \ref{tab:embed} shows, when the embedding rate is low, the detection accuracy is also low. The reason is that when the embedding rate is small, the statistical distribution of the carrier before and after steganography is small, making it more difficult to be detected. Furthermore, all the models increase remarkably with the increase of the embedding rate when the embedding rate is slow, but it is not obvious when the embedding rate is high. Because as the embedding rate increase, the sample have more clues for steganalysis leading to higher detection accuracy but it doesn't benefit more when the embedding rate is high relatively. Besides, when the embedding rate is above 20\%, the detection accuracies of our model are all above 95\%. When the embedding rate is 10\%,  the CSW model and our model are significantly better than other methods. Meanwhile, our model out-performs the CSW method by more than 11 percent  in testing Chinese samples in 10\% embedding rate. Overall, our model significantly improves the detection accuracy in low embedding rate. Generally, to avoid being easily detected, steganography algorithms often adopt low embedding rate strategy, which poses a challenge to steganalysis. Our model's excellent performance in low embedding rate makes it more practical in realistic scenario. 
In addition, we also noticed that the performance of all  models in English speech samples is better than that in Chinese speeches when the embedding rate and duration are equal in most of cases. This phenomenon may be explained by the different characteristics  of the two languages such as alphabet, grammar and phonology . Moreover, performances in different languages are all good enough to
show that our method can well adapt to different languages. 

\subsubsection{Influence of Sample Length}

\begin{figure}
\centering
\includegraphics[width=3.5in]{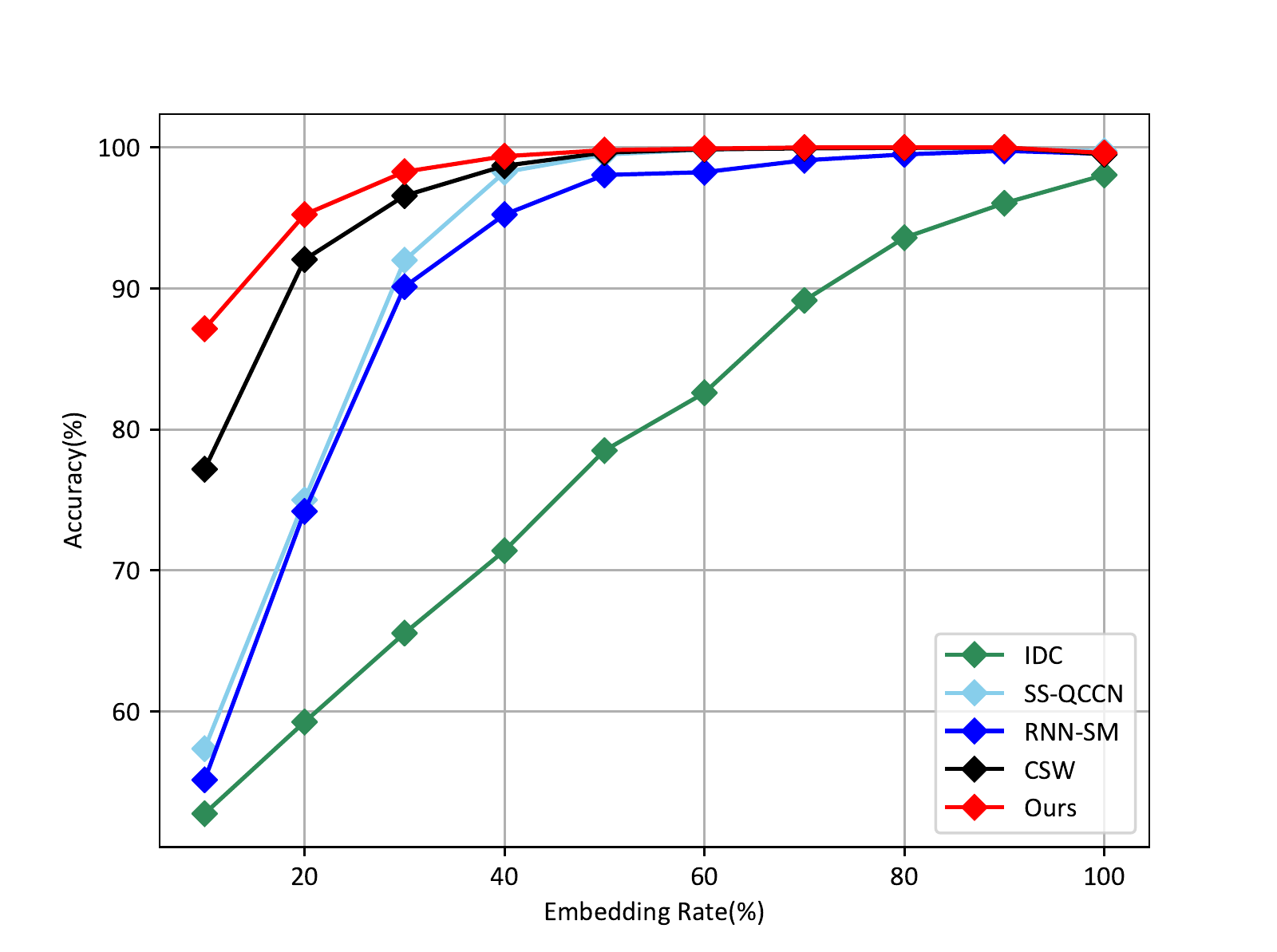}
\caption{Detection Accuracy of 100\% Embedding Rate Samples Under Different Lengths}
\label{slen}
\end{figure}

 \begin{table*}
\renewcommand\arraystretch{1.3}
  \centering
  \caption{Detection Accuracy of 100\% Embedding Rate Samples Under Different Lengths}
  \resizebox{6in}{!}{
    \begin{tabular}{c|l|cccccccccc}
    \toprule[2pt]
      \multirow{2}{*}{Language} &\multirow{2}{*}{Method}  &\multicolumn{10}{|c}{Sample Length (s)}\\
    & & 0.1 & 0.2 & 0.3 & 0.4 & 0.5 & 0.6 & 0.7 & 0.8 & 0.9 & 1 \\
    \hline
    \multirow{5}{*}{EN} & IDC \cite{Huang2011Detection} & 85.40 & 88.00 & 88.50 & 89.25 & 90.10 & 91.45 & 91.40 & 92.40 & 92.95 & 93.70 \\
     & QCCN \cite{27-Li2017Steganalysis} & 82.00 & 88.85 & 92.15 & 95.00 & 95.70 & 96.15 & 96.25 & 96.90 & 96.90 & 98.00 \\
    & RNN-SM\cite{28-Lin2018RNN} & 90.40 & 95.50 & 97.38 & 97.81 & 98.16 & 98.23 & 98.38 & 98.48 & 98.49 & 98.54 \\
      & CSW \cite{csw_yang2019realtime}& {91.59} & {95.63} & {97.40} & {97.85} & {98.21} & {98.36} & {98.40} & {98.43} & {98.49} & 98.47 \\
     & Ours &  {\textbf{92.63}} &  {\textbf{96.41}} &  {\textbf{97.85}} &  {\textbf{98.25}} &  {\textbf{98.53}} &  {\textbf{98.71}} &  {\textbf{98.80}} &  {\textbf{98.82}} &  {\textbf{98.93}} &  {\textbf{98.95}} \\
    \hline
    \hline
    \multirow{5}{*}{CH} & IDC\cite{Huang2011Detection} & 86.80 & 88.65 & 90.20 & 90.50 & 91.20 & 92.25 & 93.10 & 94.25 & 94.70 & 94.05 \\
      & QCCN\cite{27-Li2017Steganalysis} & 81.20 & 90.05 & 93.75 & 95.25 & 96.50 & 97.45 & 97.60 & 98.30 & 98.10 & 98.50 \\
       & RNN-SM\cite{28-Lin2018RNN} & 90.91 & 95.91 & 97.03 & 97.72 & 98.09 & 98.12 & 98.51 & 98.69 & 99.06 & 98.86 \\
       & CSW \cite{csw_yang2019realtime} & {91.84} & {96.12} & {97.70} & {98.32} & {98.56} & {98.40} & {98.99} & {98.80} & {99.13} & 98.95 \\
       & Ours &  {\textbf{92.33}} &  {\textbf{96.79}} &  {\textbf{98.20}} &  {\textbf{98.82}} &  {\textbf{99.11}} &  {\textbf{99.24}} &  {\textbf{99.34}} &  {\textbf{99.41}} &  {\textbf{99.42}} &  {\textbf{99.43}} \\
    \bottomrule[2pt]
    \end{tabular}%
    }
  \label{tab:sampleLen}%
\end{table*}%

The duration of voice is another factor which has great impact when detecting QIM based steganography in VoIP streams. The detection of short steganography samples is challenging. To test the performance of the proposed algorithm against different lengths of samples, we fixed the embedding rate at 100\%. As for the sample length, we tested 10 samples whose lengths are equally spaced in the range of 0.1s to 1s with step equals to 0.1s. According to the results shown in Table \ref{tab:sampleLen} and Figure \ref{slen}, we can see that when the sample length increases, the detection accuracy increases. This phenomenon is easy to explain. Longer sequence provides more observations on code-word correlations, which can therefore be modeled more accurately. Thus, the difference between the code-word correlation patterns of stego speech and cover speech is more distinct, leading to easier classification. Moreover, when the sample length is small, increasing sample length significantly benefits the accuracy. As the sample length increases, the benefit of increasing sample length diminishes. Most importantly, we can come to the conclusion that our model is better than all the previous methods when the samples are short. It means that our method can effectively detect the QIM steganography in low bit rate speech only by capturing a small segment speech stream of a monitored VoIP session, which is very important for VoIP corresponding censoring.

\begin{table}[!h]
  \renewcommand\arraystretch{1.4}
  \caption{Classfication Accuracy Under Different Length and Different Embedding Rate}
    \resizebox{3.3in}{!}{
    \begin{tabular}{|c|c|l|r|r|r|r|}
    \toprule[2pt]
    \multirow{2}[4]{*}{\textbf{Sample Length}} & \multirow{2}[4]{*}{\textbf{Language}} & \multicolumn{1}{c|}{\multirow{2}[4]{*}{\textbf{Method}}} & \multicolumn{4}{c|}{\textbf{Embedded Rates}} \\
\cline{4-7}      &   &   & 10 & 20 & 30 & 40 \\
\hline
\hline
    \multirow{10}[20]{*}{0.1s} & \multirow{5}[10]{*}{EN} & IDC \cite{Huang2011Detection} & 52.95 & 57.65 & 62.90 & 67.05 \\
      &   & QCCN \cite{27-Li2017Steganalysis}& 50.55 & 54.80 & 58.25 & 59.05 \\
      &   & RNN-SM \cite{28-Lin2018RNN} & 55.39 & 60.25 & 67.43 & 70.28 \\
      &   & CSW \cite{csw_yang2019realtime} & 55.99 & 62.44 & 67.58 & 72.36 \\
      &   & Ours & \textbf{57.02} & \textbf{64.93} & \textbf{71.04} & \textbf{76.18} \\
\cline{2-7}      & \multirow{5}[10]{*}{CH} & IDC \cite{Huang2011Detection} & 53.90 & 58.85 & 63.70 & 68.05 \\
      &   & QCCN \cite{27-Li2017Steganalysis} & 51.55 & 54.80 & 58.25 & 59.05 \\
      &   & RNN-SM \cite{28-Lin2018RNN} & 54.71 & 60.48 & 63.60 & 68.18 \\
      &   & CSW \cite{csw_yang2019realtime} & 55.20 & 61.71 & 67.15 & 72.04 \\
      &   & Ours & \textbf{56.00} & \textbf{63.32} & \textbf{69.81} & \textbf{74.84} \\
   \hline
   \hline
    \multirow{10}[20]{*}{0.3s} & \multirow{5}[10]{*}{EN} & IDC \cite{Huang2011Detection} & 54.55 & 58.15 & 63.65 & 69.50 \\
      &   & QCCN \cite{27-Li2017Steganalysis} & 53.15 & 58.25 & 62.90 & 71.45 \\
      &   & RNN-SM \cite{28-Lin2018RNN} & 59.68 & 70.05 & 77.17 & 77.27 \\
      &   & CSW \cite{csw_yang2019realtime} & 61.23 & 70.61 & 78.21 & 84.43 \\
      &   & Ours & \textbf{62.81} & \textbf{73.50} & \textbf{81.16} & \textbf{86.66} \\
\cline{2-7}      & \multirow{5}[10]{*}{CH} & IDC \cite{Huang2011Detection} & 54.50 & 60.10 & 65.70 & 70.05 \\
      &   & QCCN \cite{27-Li2017Steganalysis} & 53.15 & 58.25 & 62.90 & 71.45 \\
      &   & RNN-SM \cite{28-Lin2018RNN} & 57.61 & 66.81 & 74.60 & 80.08 \\
      &   & CSW \cite{csw_yang2019realtime} & 59.66 & 70.02 & 77.87 & 82.75 \\
      &   & Ours & \textbf{60.73} & \textbf{71.76} & \textbf{85.05} & \textbf{85.11} \\
   \hline
  \hline
    \multirow{10}[20]{*}{0.5s} & \multirow{5}[10]{*}{EN} & IDC\cite{Huang2011Detection} & 51.85 & 59.85 & 64.05 & 72.30 \\
      &   & QCCN \cite{27-Li2017Steganalysis} & 53.65 & 61.85 & 67.35 & 75.20 \\
      &   & RNN-SM \cite{28-Lin2018RNN} & 62.46 & 72.45 & 80.38 & 86.22 \\
      &   & CSW \cite{csw_yang2019realtime} & 64.33 & 75.72 & 83.80 & 89.49 \\
      &   & Ours & \textbf{66.57} & \textbf{79.23} & \textbf{85.83} & \textbf{91.26} \\
\cline{2-7}      & \multirow{5}[10]{*}{CH} & IDC \cite{Huang2011Detection} & 56.55 & 59.60 & 65.45 & 70.15 \\
      &   & QCCN \cite{27-Li2017Steganalysis} & 53.65 & 61.85 & 67.35 & 75.20 \\
      &   & RNN-SM \cite{28-Lin2018RNN} & 71.43 & 71.29 & 78.42 & 84.39 \\
      &   & CSW \cite{csw_yang2019realtime} & 62.45 & 74.21 & 82.19 & 86.83 \\
      &   & Ours & \textbf{64.03} & \textbf{76.21} & \textbf{84.09} & \textbf{90.03} \\
    \bottomrule[2pt] 
    \end{tabular}
    }
  \label{tab:result}%
\end{table}%

\subsection{Time Efficiency of Different Model}
\begin{figure}
\centering
\includegraphics[width=3.7in]{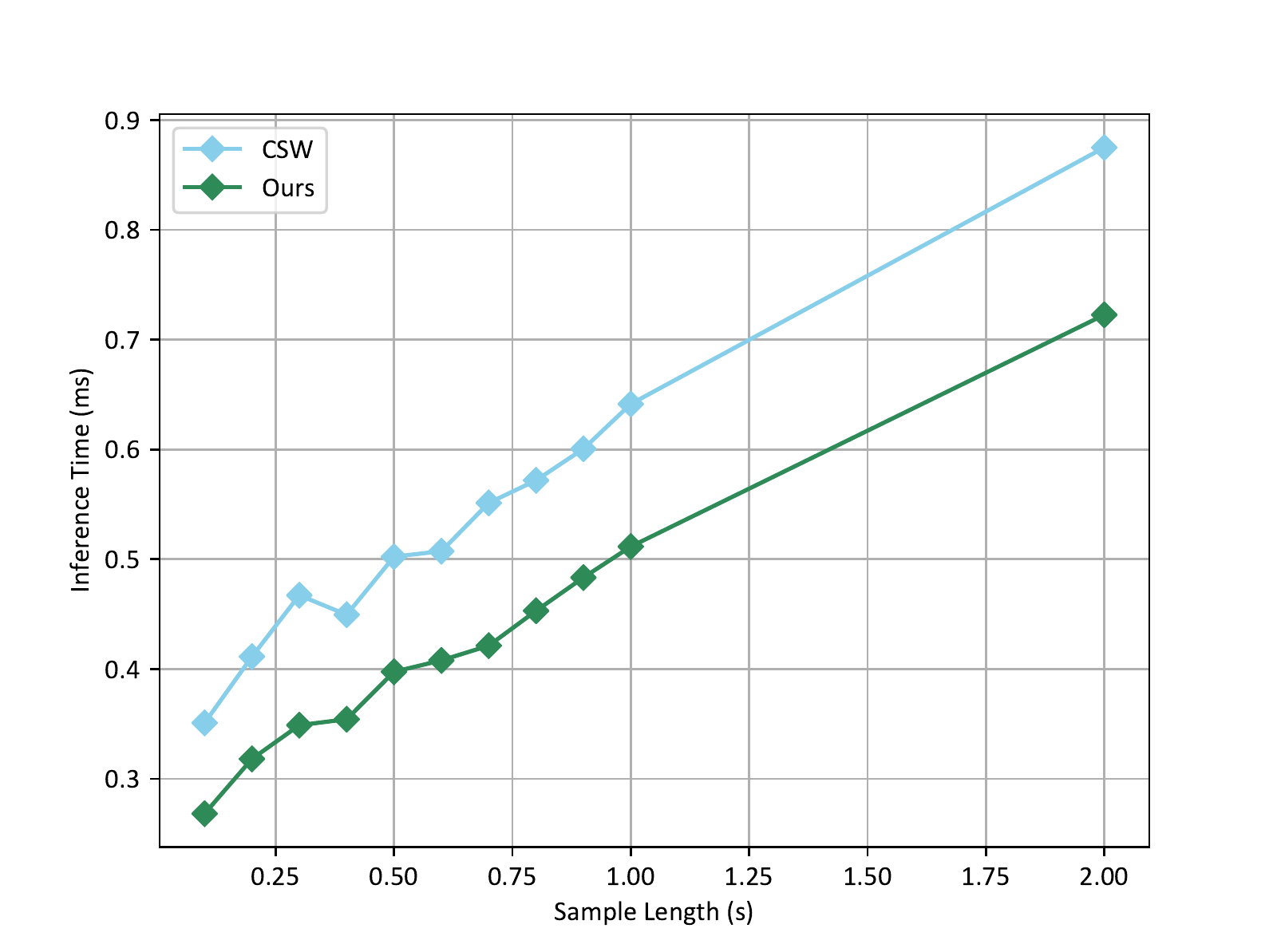}
\caption{Time Efficiency of Different Models}
\label{timef}
\end{figure}

Time efficiency is also an important factor in determining whether a model can actually be applied to an online scene. Yang $et\ al.$ \cite{csw_yang2019realtime} have demonstrated in their article that CSW significantly outperforms other methods in terms of time efficiency. Thus, our experiments only compare the time efficiency of our model with the CSW method. The results are shown in the table \ref{tab:time} and figure \ref{timef}. Obviously, from the table, our model perform significantly better than the CSW method in various sample lengths. Especially, when the sample lenght is short, such as 0.1s, inference time of proposed model is only 2/3 of the CSW method. In addition, we also noticed that because the CSW model uses multi-channel convolution to extract features and our model used share convolution with single convolution path, parameters of proposed model have been significantly reduced. For instance, when the sample length is 1s, parameters of proposed model are also 1/4 of the CSW method, which make it more easy to apply at real sense .

\begin{table}[htbp]
  \centering
\renewcommand\arraystretch{1.4}
  \caption{Paramters of Different Model}
    \begin{tabular}{|l|r|}
    \hline
    Model & \multicolumn{1}{l|}{Total Parameters} \\
\hline    CSW \cite{csw_yang2019realtime}& 567,041 \\
\hline    Ours & \textbf{157,825} \\
\hline
    \end{tabular}%
  \label{tab:param}%
\end{table}%

\subsection{Discussion of Model variants}
In this part, we try to investigate the function  of different parts in the proposed model by comparing it with its several variants. Performance of different variants are shown in table \ref{tab:modelv}.

First, comparing \#0, \#1, \#2 and \#3, we can see that shortcut connections in different aspects are beneficial to steganalysis. It is easy to explain that steganography will change the structure of speech in different aspects, and features from different shortcut connections provide abundant information for detection. 
\begin{table}[!h]
\renewcommand\arraystretch{1.3}
  \centering
  \caption{The detection accuracy of various models}
  \resizebox{3.5in}{!}{
     \begin{tabular}{c|l|c}
    \toprule[1pt]
    \multicolumn{1}{l|}{\textit{\textbf{ Index}}} & \multicolumn{1}{l|}{\textit{\textbf{Network Description}}} & \multicolumn{1}{l}{\textit{\textbf{Accuracy}}} \\
    \hline
    \textbf{\#0} & The proposed model &\textbf{87.14} \\
    \hline
    \textbf{\#1} & Remove path 1 & 86.97 \\
   \hline   
    \textbf{\#2} & Remove path 2 & 86.76 \\
    \hline
    \textbf{\#3} & Remove 1 and 2 & 86.82 \\
    \hline
    \textbf{\#4} & Replace attention with max pooling & 79.63 \\
    \hline
    \textbf{\#5} & Reduce convolution block to 2 & 82.06 \\
    \hline
    \textbf{\#6} & Add convolution block to 4 & 85.59 \\
    \bottomrule[2pt]
    \end{tabular}%
    }
  \label{tab:modelv}%
\end{table}%
Meanwhile, the attention mechanism is used in our model to select important information as well as to reduce dimensions. However, pooling \cite{pooling-Krizhevsky2012ImageNet} is the most common way to reduce the dimension of features which has a similar function attention mechanism. Hence, we replaced the attention mechanism with max pooling operation to show the effectiveness of attention mechanisms. It is obvious that giving different weights to different vectors are helpful to our model when \#0 is compared with \#4. Moreover, models in \#0, \#6 and \#7 shows that three convolution blocks are proper in our experiments. In general, more features of the
input data can be captured by a deeper network and the difference between model \#0 and \#6 proves that. However, performance of \#0 and \#7 demonstrate it doesn't mean that the deeper the network is, the better  the model performance will be, since deeper networks may result in over-fitting and vanishing gradient problems.
\begin{table*}[!htbp]
\renewcommand\arraystretch{1.9}
  \centering
  \caption{Detection Accuracy of 100\% Embedding Rate Samples Under Different Lengths}
  \resizebox{6in}{!}{
    \begin{tabular}{|c|l|r|r|r|r|r|r|r|r|r|r|r|}
     \toprule[2pt]
    \multirow{2}[0]{*}{Method} & \multicolumn{1}{|c|}{\multirow{2}[0]{*}{Metric}} & \multicolumn{11}{|c|}{Sample Length (s)} \\
     \cline{3-13} 
      &   & 0.1 & 0.2 & 0.3 & 0.4 & 0.5 & 0.6 & 0.7 & 0.8 & 0.9 & 1 & 2 \\
\hline
    \multirow{2}[0]{*}{CSW} & Mean (ms) & 0.3509 & 0.4113 & 0.4671 & 0.4494 & 0.5023 & 0.5073 & 0.5512 & 0.5718 & 0.6005 & 0.6413 & 0.8749 \\
    \cline{2-13} 
      & Std & 0.1490 & 0.1957 & 0.2033 & 0.1637 & 0.1829 & 0.1446 & 0.1806 & 0.2000 & 0.2059 & 0.2160 & 0.2913 \\
\hline
    \multirow{2}[0]{*}{Ours} & Mean (ms) & \textbf{0.2683} & \textbf{0.3181} & \textbf{0.3488} &\textbf{ 0.3542 }& \textbf{0.3975} &\textbf{ 0.4078} & \textbf{0.4212} & \textbf{0.4530 }&\textbf{ 0.4833} & \textbf{0.5115} & \textbf{0.7225} \\
     \cline{2-13} 
      & Std & 0.1322 & 0.1613 & 0.1581 & 0.1594 & 0.1636 & 0.1406 & 0.1292 & 0.1492 & 0.1609 & 0.1563 & 0.2373 \\
\bottomrule[2pt] 
    \end{tabular}%
    }
    
  \label{tab:time}%
\end{table*}%

\section{Conclusions}
VoIP is a very popular streaming media for steganography. Detecting short and low embeded QIM steganography samples in VoIP stream remains an unsolved challenge in real circumstances. Potential VoIP-based covert communications based on QIM steganography pose a great threat to the security of cyberspace. Previous methods in steganalysis of QIM based steganography always pay much attention to the correlations in inter-frames and intra-frames but ignore the hierarchical structure in speech frames. In this paper, motivated by the complex multi-scale structure which appears in speech, we proposed  hierarchical representative network to address this steganalysis problem in VoIP streams. In our model, CNN is stacked to model hierarchical structure in speech and attention mechanisms are applied to select import information. Experiments demonstrate that our model is effective and can achieve state-of-the-art result. Besides,our model needs less computation and has higher time efficiency to be applied to real online services. Although our model performs well enough, detection accuracy in low embedded rate still needs improvement.




%

\section*{Acknowledgment}


\ifCLASSOPTIONcaptionsoff
  \newpage
\fi





\bibliographystyle{./IEEEtran}
\bibliography{sample-bibliography}

\end{document}